\documentclass[pra,twocolumn,showpacs,superscriptaddress]{revtex4-1}

\usepackage{graphicx}
\usepackage{subfigure}
\usepackage[english]{babel}
\usepackage{amsfonts}
\usepackage{amsmath}
\usepackage{latexsym}
\usepackage{bm}
\usepackage{dcolumn}
\usepackage{bm}
\usepackage{rotating}
\usepackage[usenames]{color}
\usepackage{mathcomp}

\usepackage{epstopdf}

\def\bM{{\bf M}}

\def\bA{{\bf A}}
\def\cE{{\cal E}}

\begin{document}
\title{Density of states in an optical speckle potential}

\author{G. M. Falco}
\affiliation{Institut f\"ur Theoretische Physik, Universit\"at zu
K\"oln, Z\"ulpicher Str. 77, D-50937 K\"oln, Germany}

\author{A. A. Fedorenko}
\affiliation{Institut f\"ur Theoretische Physik, Universit\"at zu
K\"oln, Z\"ulpicher Str. 77, D-50937 K\"oln, Germany}
\affiliation{CNRS-Laboratoire de Physique de l'Ecole Normale Sup{\'e}rieure de Lyon, 46,
All{\'e}e d'Italie, 69007 Lyon, France}
\author{J. Giacomelli}
\affiliation{Sace, Piazza Poli 37/42, 00187 Roma, Italy}

\author{M. Modugno}
\altaffiliation{also at Dipartimento di Fisica e Astronomia \& LENS, Universit\`a di Firenze, 50019 Sesto Fiorentino, Italy}
\affiliation{Department of Theoretical Physics and History of Science,
UPV-EHU, 48080 Bilbao, Spain}

\affiliation{IKERBASQUE, Basque Foundation for Science, 48011 Bilbao, Spain}

\date{November 8, 2010}

\begin{abstract}

We study the single particle density of states of a one-dimensional
speckle potential, which is correlated and non-Gaussian.
We consider both the repulsive and the attractive cases.
The system is controlled by a single dimensionless parameter determined
by the mass of the particle,
the correlation length and the average intensity of the field.
Depending on the value of this parameter, the system exhibits
different regimes, characterized by the localization properties
of the eigenfunctions. We calculate the corresponding density of states
using the statistical properties of the speckle potential. We find good
agreement with the results of numerical simulations.

\end{abstract}
\pacs{37.10.Jk, 42.30.Ms, 03.75.-b \hspace{22mm}
Published: Phys. Rev. A 82, 053405 (2010). }

\maketitle


\section{Introduction}

Speckles are high-contrast fine-scale granular patterns occurring whenever
radiation is scattered from a surface characterized by some roughness on the
scale of the radiation wavelength.
Originally discovered for laser light in the early 1960s, the speckle phenomenon
plays an important role not only in optics, but also in other fields, where it is
used for several applications, including, for example, radar and ultrasound medical
imagery \cite{goodman}.
In recent years, optical speckles have been employed in combination with cold
atoms \cite{boiron} and especially Bose-Einstein condensates (BECs) \cite{lye05}
in order to investigate the behavior of matter waves in the presence of
disordered potentials \cite{lye05,clement05,fort05,schulte05}. Many interesting
features of BECs in one-dimensional (1D) speckle potentials have been addressed
from both the experimental and the theoretical sides. These include classical
localization
and fragmentation effects, frequency shifts and damping of collective
excitations, and inhibition of transport properties
\cite{lye05,clement05,fort05,schulte05,modugno06,palencia06,clement06,Lugan07,%
sanchez07,chen08,billy08,roati08,sanchez08,dries10},
and have culminated with the observation of Anderson localization for a
BEC \cite{sanchez07,billy08,roati08}. Also, the
superfluid-insulator transition
\cite{Fisher89,Zhou06,Glyde07,Falco09b,Guraries09,Fontanesi10,Altman10,Deissler}
and the transport of coherent matter waves
have been recently
investigated from the theoretical point of view for speckles
in higher dimensions \cite{kuhn05,kuhn07,henseler08,pilati}.

Despite this intense research activity, the properties of the single particle
spectrum of the speckle potential have been addressed only partially, even in
the 1D case. In \cite{Lugan07} it has been argued that for low energies the
density of states (DOS) of \textit{blue-detuned} repulsive speckles is characterized
by a Lifshitz tail \cite{Pastur88}, whereas the presence -- for higher energies --
of an effective mobility edge, has been discussed in \cite{sanchez07}.
In addition, though there is a vast literature about random 1D
systems \cite{Pastur88,Kramer93}, most of the general theorems apply strictly
to the case of uncorrelated disorder,
whereas the speckles, being a correlated disordered potential, deserve an
explicit treatment.

In this article we discuss the properties of DOS for a 1D
speckle potential, by comparing analytical predictions, based on the
statistical properties of the speckles, with explicit calculations for
the spectrum of numerically generated speckle patterns.
We consider
both \textit{blue-detuned} (repulsive) and \textit{red-detuned} (attractive) cases.
The speckles are characterized by their
intensity $I_{0}$ and a correlation length $\xi$, which represents
the length scale over which the modification of the potential occurs. Therefore,
for dimensional reasons, the single-particle properties are determined by a
single dimensionless parameter $s=\pm{2m\xi^2I_0}/{\hbar^2}$,
with $m$ being the mass of the particle. In our conventions, the plus
and the minus signs refer to the blue-detuned and red-detuned speckle, respectively.

In the presence of disorder, the asymptotic behavior of the DOS
near the edges of the spectrum is dominated by rare large
fluctuations of the potential.
For the red-detuned potential, the low-energy tail of the DOS
originates from the deep wells at negative energies.
The latter is closely related to the distribution of intensity maxima in
the speckle patterns.  In the case of a blue-detuned speckle
the spectrum is bounded from below. The DOS near the edge results
from the large regions with very small intensity.
For both cases we analyze the behavior of the DOS throughout
the $s$ range, from the semiclassical limit $|s|\to\infty$, down to the quantum
regime $|s|<1$, where smoothing effects become important.

The article is organized as follows. In Secs.~\ref{sec:speckle} and~\ref{sec:Weinrib}
we introduce the speckle potential and its statistical properties.
Then in Sec.~\ref{sec:red} we discuss the DOS for the
attractive \textit{red-detuned} speckles, for different regimes of $s$.
The case of the \textit{blue-detuned} repulsive speckles is considered in
Sec.~\ref{sec:blue}. Our results are summarized in~Sec.~\ref{sec:sum}.

\section{The speckle potential}
\label{sec:speckle}

The electric field $\cE(x)$ created by a laser speckle on a 1D
Euclidean line at the image plane is, to a very good approximation,
a realization of a complex Gaussian
variable. In general,  $\cE(x)$ can be taken as statistically isotropic in the
complex $\cE$ plane and statistically homogeneous and isotropic on the $x$ line.
Thus,  $\langle \cE(x)\rangle$
and $\langle \cE(x)\cE(y)\rangle$ vanish while the autocorrelation function
is given by \cite{goodman,huntley}
\begin{equation}
\label{eq:sinc_one_d}
\langle \cE^{\ast}(x)\cE(y)\rangle\equiv G(x-y)=
I_0 ~{\rm sinc}\left[D(x-y)\right],
\end{equation}
where ${\rm sinc}(t)\equiv\sin{\pi t}/(\pi t)$
and $D$ is the aperture width.
The probability distribution  of intensity $I(x)\equiv |\cE(x)|^2$
across a speckle pattern follows a negative-exponential
(or Rayleigh) distribution \cite{huntley,modugno06},
\begin{equation}
\label{eq:ray}
P[I]=\exp{[-I/I_0]}/I_0
\end{equation}
where  $I_{0}=\langle I\rangle$ is the mean intensity while the most probable
intensity is zero.
The intensity autocorrelation function is given by
$\langle
I(x)I(0)\rangle =I_0^2\left[1+ {\rm sinc}^2(D x)\right]$.
The (auto) correlation length $\xi$ of the speckle potential
is defined through the equation ${\rm sinc}^2(D{\xi}/{2})=1/2$,
which gives the width at the half
maximum of the autocorrelation function. It is related to the aperture width by
$\xi=0.88/D$.

When the speckle electric field is shined on a sample of atoms, this
results in a disordered potential felt by the atoms, that is proportional
to the local intensity of the speckles, $V(x)=\alpha I(x)$ with
$I(x)\equiv |\cE(x)|^2$. In general, if the wavelength of the light is
far detuned from the atomic resonance, the constant $\alpha$ is proportional to
the inverse of the detuning $\Delta$, and  can be either positive or
negative \cite{fallani08}.
This corresponds to a disordered potential bounded from below and composed by a
series of barriers ($\Delta>0$, \textit{blue detuning}) or bounded from above
and made of potential wells ($\Delta<0$, \textit{red detuning}) .
In what follows, we include  $|\alpha|$ in redefinition of $I_0$, which
is positive as well as $I(x)$. Then we use $\alpha=+1$ and
$\alpha=-1$ for the blue-detuned and red-detuned speckles, respectively.

The quantum properties of a particle of mass $m$ in the speckle
potential are determined by the solutions of the Schr{\"o}dinger equation,
\begin{equation}
\label{eq:schroe}
\left[-\frac{\hbar^2}{2m}\nabla^2_{x} +\alpha I(x)\right]\psi(x)=E{\psi(x)},
\end{equation}
with $\psi(x)$ being the particle wave function.
Note that there are two relevant length scales in the
problem~\cite{Falco08}. The first one is
the correlation length $\xi$ of the disorder which introduces the corresponding
energy scale,
\begin{equation}\label{eq:en_sc_xi}
E_{\xi}=\hbar^2/2m\xi^2.
\end{equation}
Additionally to this length scale, the particle of mass $m$ moving
in the random potential with typical strength $I_0$
introduces a new length scale $B$ defined by
$$I_0={\hbar^2}/{2 m B^2}.$$
By comparing $I_0$ with $E_{\xi}$ the dimensionless
parameter $s$ can be written as
\begin{equation}
\label{eq:s}
s=\alpha \left({\xi}/{B}\right)^2.
\end{equation}
For large values of $|s|$ the disorder is strongly correlated while
for $|s|\rightarrow 0$ the limit of
uncorrelated disorder is approached.
Expressing the intensity of the speckle potential in units of $I_0$
so that  $\langle \tilde{I}\rangle=1$ and the length in units of $\xi$,
we can rewrite the  Hamiltonian in a dimensionless form as
$\tilde{H} = -\nabla^2_{\tilde{x}} + s \tilde{I}(\tilde{x})$ with a single
parameter~$s$.

\begin{figure}
\includegraphics[width=0.97\columnwidth]{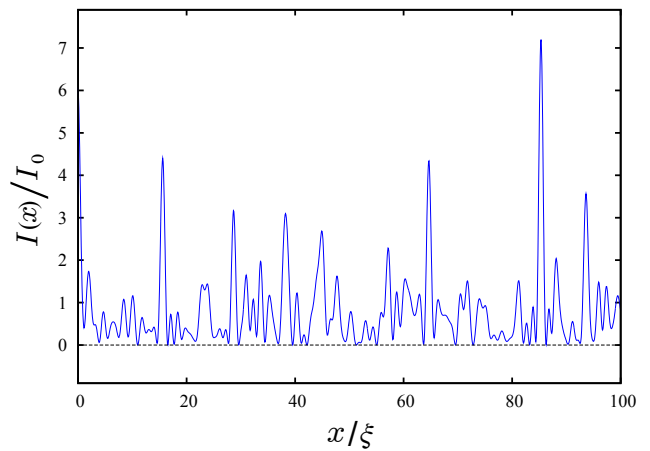}
\caption{(Color online) The typical 1D (repulsive) blue-detuned speckle profile.
The intensity is given
in units of $I_0$ and the coordinate in units of $\xi$. }
  \label{fig:speckle}
\end{figure}

Once the electric field correlation function~(\ref{eq:sinc_one_d}) is known,
the statistical properties of the intensity
are completely determined.
This is discussed in the next section, where we consider the statistics
of extrema of this correlated and non-Gaussian potential,  relevant
for the motion of a particle in a random landscape \cite{ledoussal03}.
By using this description,  we derive
the DOS and compare the analytical predictions  with the
spectrum computed numerically for a randomly generated speckle pattern.

To generate the speckle potential profile we use the mathematical counterpart
of the mechanism that gives rise to optical speckles. The constructive
or destructive
interference of randomly phased elementary components of the radiation
field caused by the roughness of the scattering surface
occurs also in the discrete Fourier transform of a sample function of a
random process \cite{goodman}. This effect can be exploited to
generate numerically a speckle pattern with the statistical properties
discussed earlier \cite{huntley,modugno06}. Then, the spectrum of the system
can be computed numerically by mapping the stationary
Schr\"odinger equation $H\psi=E\psi$ on a discretized grid \cite{numerical}.
All the results presented in this article have been obtained for a system of
length $L=600\xi $ and are averaged over a number of realizations ranging from
$10^{2}$ to $10^{3}$.

\section{Statistical analysis of the speckle potential}
{\label{sec:Weinrib}}

The typical shape of the potential profile created by a blue-detuned
speckle pattern is shown in Fig.~\ref{fig:speckle}.
The potential profile can be seen as series of
peaks (maxima) statistically distributed around the typical value
of the order of $I_0$. The peaks are
separated by valleys of typical width $\xi$ with minima situated most probably
near the lower-boundary of the potential.
For  the red-detuned case the picture is analogous, with the roles
of maxima and minima exchanged. In other words, the statistics of
the potential minima for the red-detuned case is that of the
maxima of the blue-detuned one.

We are interested in the density of minima $n_{\rm min}\left(I'\right)$
and the density of maxima $n_{\rm max}\left(I'\right)$, at specified intensity
$I'$, in the spatial distribution $I(x)$ of intensities along the image line.
Using the method of Weinrib and Halperin~\cite{Weinrib82}, the number of minima
or maxima can be calculated from the integrated joint probability,
\[
N_{\rm min, max}={\int}dx ~\delta\left({I_x}\right)
{\Big|}{I_{xx}}{\Big|}\Theta\left({\pm I_{xx}}\right),
\]
where the plus (minus) sign refers to minima (maxima) and we have used the conventions
${\partial I}/{\partial x}\equiv I_x$ and
${\partial^2 I}/{\partial x^2}\equiv I_{xx}$. It follows that the density
of minima (maxima)
per unit length and intensity
can be written as
\begin{equation}
\label{eq:dens_min_one_d}
n_{\rm min, max}\left(I'\right)=\langle \delta\left({I_x}\right)\delta\left(I'-I\right)
{|}{I_{xx}}{|}\Theta\left({\pm I_{xx}}\right)\rangle.
\end{equation}
In order to evaluate the statistical average of Eq.~(\ref{eq:dens_min_one_d}),
we need to calculate the joint probability
of $I$, $I_x$ and $I_{xx}$ (all at the same point) starting from the joint
probability of the Gaussian
random variables $\cE$, $\cE_x$, and $\cE_{xx}$. Because $\cE(x)$ is statistically
homogeneous along the $x$ coordinate,
the density in Eq.~(\ref{eq:dens_min_one_d}) will be independent of position.

\begin{figure}
\includegraphics[width=0.9\columnwidth]{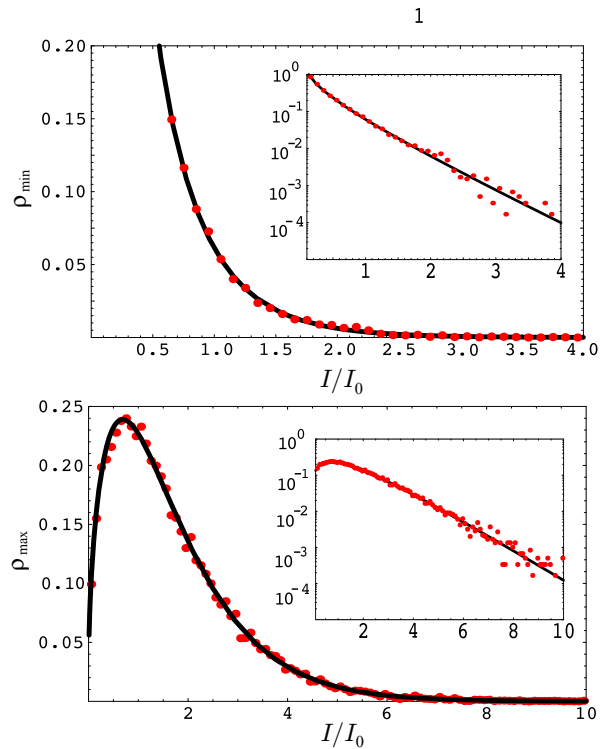}
  \caption{(Color online) Density $\rho(I/I_0)$ of minima (top)
  and maxima (bottom) for a blue-detuned speckle (in log scale in the insets).
  The solid lines represent the analytical prediction of Eq.~(\ref{eq:rhomin}),
  while the dots have    been obtained
  by averaging   over $100$  numerical speckle realizations.
}
  \label{fig:min_max}
\end{figure}

The  statistics of a set of Gaussian variables is completely determined
by the two-point correlations $G(x-y)$ between them.
The nonzero correlators evaluated at the same point are calculated by considering
the derivatives of $G(x-y)$. One  finds
$\langle |\cE|^2 \rangle  =G(0),
\langle \cE_{xx} \cE^{\ast} \rangle =-\langle |\cE_x|^2\rangle=G_{xx}(0),
\langle |\cE_{xx}|^2 \rangle=G_{xxxx}(0)$ with
\begin{align}
\label{eq:correlations_one_d_values}
G(0) &= I_0,\\
G_{xx}(0) & =-{I_0}/{\tilde{\xi}^2},\\
G_{xxxx}(0) &={I_0}/(\kappa~\tilde{\xi}^4),
\end{align}
where we have introduced the length $\tilde{\xi}\equiv \sqrt{3}/(\pi D)$
for convenience. The dimensionless parameter
\begin{equation}
\kappa= \frac{G_{xx}^2(0)}{G(0)G_{xxxx}(0)}
\end{equation}
is restricted in one dimension to the range $0\leq\kappa\leq 1$.
It depends only on the the shape of the
correlation function of the electric field but not on the length
scale  $\xi$. For the correlator of the laser speckle given by
Eq.~(\ref{eq:sinc_one_d}), we have $\kappa=5/9$.

On dimensional grounds, it is clear that the density of minima
(maxima) of Eq.~(\ref{eq:dens_min_one_d}) scales as
\begin{equation}
n_{\rm min,max}(I,\tilde\xi,\kappa)=\left({1}/{(I_0 \tilde\xi})\right)
\rho_{\rm min,max}\left({I}/{I_0},\kappa\right).
\end{equation}
Evaluating the statistical average in Eq.~(\ref{eq:dens_min_one_d})
and rewriting in terms of dimensionless intensity measured in units of $I_0$,
we obtain
\begin{align}
\label{eq:rhomin}
\rho_{\rm min,max}&\left({\tilde{I}},\kappa\right)=\frac{e^{-\frac{\tilde{I}}{1-\kappa}}}{4 \pi^{3/2}}
\frac{1}{\tilde{I}^{1/2}\sqrt{\kappa^{-1}-1}}\nonumber \\
\times&\int_{-\infty}^{\infty}d \theta_x ~f(\tilde{I},\theta_x)
\int_{0}^{\infty}d \tilde{I}_{xx}~
g(\tilde{I},\theta_x,\tilde{I}_{xx}),
\end{align}
where $f=\exp\left[-\tilde{I}\theta_x^2
\left(
\theta_x^2-3+\kappa^{-1}
\right)/\left(\kappa^{-1}-1\right)
\right]$, \\
$g=\tilde{I}_{xx}~
\exp{\left\{-\tilde{I}_{xx}\left[\left({\tilde{I}_{xx}}/{4~\tilde{I}}\right)
\mp{\left(\theta_x^2-1\right)}\right]
/\left(\kappa^{-1}-1\right)\right\}}$,
and the plus (minus) sign refers to the minima (maxima).
The details of the derivation are given in the Appendix.
The densities~(\ref{eq:rhomin}) are plotted
in Fig.~\ref{fig:min_max}.
The agreement with the numerical simulation is excellent.

The asymptotic behavior of the integrals in Eq.~(\ref{eq:rhomin}) can be also
evaluated analytically. This is useful because
the distribution of the deep negative-energy wells of the red-detuned speckle
potential is related to the large $I$
asymptotic tail of the maxima distribution of the blue-detuned speckle.
This latter behaves for $\kappa=5/9$ as
\begin{eqnarray}
{\rho_{\rm max}}(\tilde{I})&\approx & \frac{e^{-2 \tilde{I}}}{2\pi}
\left(2+\frac{e^{\tilde{I}}
\left(-1+2 \tilde{I}\right)\sqrt{\pi}~{\rm erf}
[\tilde{I}]}{\sqrt{\tilde{I}}}\right) \nonumber \\
&\approx &
{\sqrt{\frac{\tilde{I}}{\pi}}} e^{-\tilde{I}}.
\end{eqnarray}
Near the boundary of the blue-detuned speckle potential
the density of minima diverges as
\begin{equation}
\label{eq:min}
{\rho_{\rm min}}(\tilde{I}) \approx {1}/{2\sqrt{\pi \tilde{I}}}.
\end{equation}
However, in the thermodynamic limit, the DOS of a particle in disordered
potential bounded from below
is expected to vanish at the boundary.
Hence, the correlation between the DOS and the number
of minima has to be rather weak.

\begin{figure}
\includegraphics[width=0.95\columnwidth]{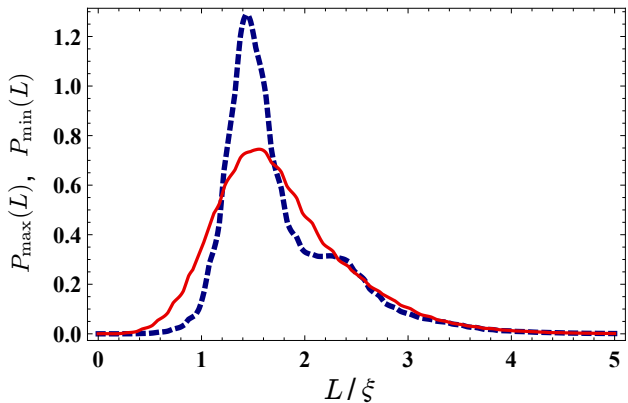}
\caption{(Color online) The normalized distributions of distance measured in
units of the autocorrelation
 length $\xi$
between neighboring maxima (dotted line) and
between neighboring minima (solid line).
}
  \label{fig:distance}
\end{figure}
More likely, the DOS near the boundary of the spectrum is associated with
the distribution of distance between
neighboring minima $P_{\mathrm{min}}(L)$ and neighboring maxima
$P_{\mathrm{max}}(L)$.
It is known that the presence of Lifshitz tails in the  DOS for
a bounded-from-below potential is related to the existence of large
regions free of disorder for the binary distribution or large regions with
negligible potential for continuous distributions.
In the case of \textit{blue-detunded } speckle potential, the zero
intensity is the most probable so that one would expect that such regions are
not so rare.
Some idea about their statistical properties can be gained through
the distributions of distances between minima and
maxima, which can be computed numerically.
The normalized  distributions of the distance between maxima and between
minima are shown in Fig.~\ref{fig:distance}. The average distance for the
both distribution is $\langle L\rangle=1.78~\xi$. This value can be easily obtained
by integrating the density in Eq.~(\ref{eq:rhomin}) over all the intensities
and by assuming that the points of minima (or maxima) are homogeneously distributed.
The right tails of both distributions can be well fitted in the
window $L/\xi \in [3,5.5]$ by simple exponentials:
\begin{eqnarray}
P_{max}(L) &\approx& 2.081 e^{-1.933(L/\xi)}, \\
P_{min}(L) &\approx& 3.385 e^{-2.036 (L/\xi)}.
\end{eqnarray}
The exponentially small probability of existence of a region of
size $L$ with relatively  small intensity accounts for the Lifshitz tail
in the DOS of the blue detuned speckle potential.

\section{DOS: red-detuned speckle}
\label{sec:red}

We  consider first the DOS for a particle in the
red-detuned speckle potential, which corresponds to $s<0$.
We discuss separately three different regimes,
ranging from the semiclassical limit, for $s\rightarrow -\infty$, down to
the quantum regime, when $|s|\le1$.

\begin{figure}
\includegraphics[width=0.9\columnwidth]{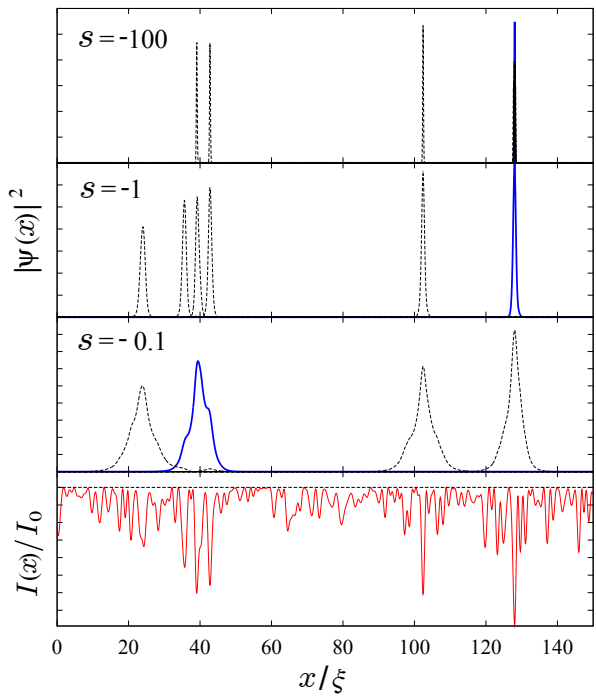}
  \caption{(Color online) Probability density distribution of the lowest lying
eigenstates of a \textit{red-detuned} speckle pattern (bottom panel)
for three different intensities corresponding to $s=-0.1,\ -1,\ -100$.
Only a window of length $150\xi$ of the whole system is shown.
The figure shows the ground state (blue solid line) and several first
excited states (black dashed line). Note that, as the speckle intensity changes,
the position of the eigenstates may change and some may move in or out
from the selected window.
}
  \label{fig:eigenstates}
\end{figure}

By varying the parameter $s$ the characteristic extension of the localized states
differs considerably.
In Fig. \ref{fig:eigenstates} we present the lowest-lying
eigenstates for three different intensities of the same speckle profile,
corresponding to $s=-0.1,\ -1,\ -100$. This picture shows that
for shallow potentials (the \textit{quantum} regime) the eigenstates extend
over several potential wells.
The lowest-lying eigenstates localize inside deep wells. However, the ground
state is not necessarily localized  in the deepest
well since the width of the well plays also a crucial role.
As the speckle intensity is increased,
the eigenstate width shrinks and their positions may change. Eventually, for
large enough $|s|$, the eigenstates tend to stack up as bound states inside
isolated wells, with the ground state being the lowest eigenstate of
the deepest well (\textit{semiclassical} regime).

\subsection{Semiclassical regime}

\begin{figure}
\includegraphics[width=0.89\columnwidth]{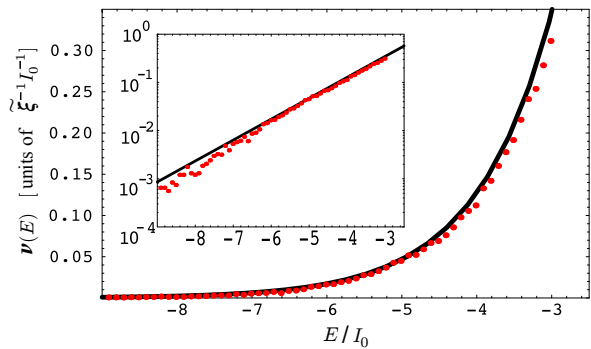}
 \caption{(Color online) Large negative-energy DOS expressed in
  units of $(\tilde\xi I_0)^{-1}$ in the presence of a red-detuned speckle
  potential  with $s=-2000$  as a function of the rescaled energy $E/I_0$
  in linear scale    and in logarithmic   scale ({inset}).
  The dots represent the numerical data (averaged over $100$ realizations),
  while the  solid line indicates the analytical
  semiclassical curve~(\ref{eq:TF_d_less}).
        }
   \label{fig:semiclassical}
\end{figure}

In the limit $s\rightarrow -\infty$
the low-energy tail of the DOS is determined by deep isolated
states occupying single wells
of typical size $\xi$. Therefore, for $E<0$,
the DOS can be calculated in
the semiclassical approximation~\cite{Kane63}
\begin{equation}
\label{eq:TF}
\nu_{|s|\rightarrow\infty} (E)=\frac{1}{I_0}\int_{-\infty}^{{E}} \nu_0(E-I) ~e^{-{{|I|}}/I_0}~d I,
\end{equation}
where $\nu_0(z)={\sqrt{m}}/{\sqrt{2}\pi\hbar\sqrt{z}}$
is the density of state of the continuum in one dimension.
Integrating over the intensity in Eq.~(\ref{eq:TF}) we find
\begin{equation}
\label{eq:TF_d_less}\nu_{|s|\rightarrow\infty} (E)
=\sqrt{\frac{m}{2 \pi\hbar^{2}I_{0}}}e^{-|E|/I_0}.
\end{equation}
The DOS in the negative-energy tail predicted by
Eq.~(\ref{eq:TF_d_less})
is in excellent agreement with the curve obtained in the numerical simulation
as shown in Fig.~\ref{fig:semiclassical}.
Note that this result differs from the DOS of a long-range correlated
Gaussian potential
that,  sufficiently deep in the tail, has the
form  $\nu(E) \sim \exp(- \mathrm{const} |E|^2)$ \cite{Pastur88}.

The semiclassical result of Eq.~(\ref{eq:TF_d_less}) can be also understood
in terms of the density
of minima
of the red-detuned speckle potential. Since the latter is the blue-detuned
speckle potential taken with negative sign, the distribution of minima
is nothing but $n_{\rm max}$, namely,
the distribution of maxima of Eq.~(\ref{eq:dens_min_one_d})
studied in Sec.~\ref{sec:Weinrib}.
We assume  that each deep state occupies an isolated single well associated
with the corresponding minimum of the potential.
This approximation is not true in general but is valid
for deep enough states. The potential inside of the well can be
completely described by derivatives taken at the
bottom of the well, denoted by
$\{I\} := \{I, I_{x}=0,I_{xx},I_{xxx},...\}$. Each well
is characterized by  $M\left[\{I\}\right]$ bound states
with energies  $E_n$, $n=1, \dots, M$,
which can be found from the solution of the Schr\"odinger
equation for the particle in the
potential $I(x)= -(I + \frac12I_{xx} x^2 + \frac16 I_{xxx} x^3 +...)$.

The DOS can be then calculated by summing over all levels in each well
and over all wells as follows
\begin{equation}
\label{eq:harm}
\nu(E)=\int d\{I\}\!\! \sum_{n=0}^{M\left[\{I\}\right]} n_{\rm min}^{\rm red}\left(\{I\}\right)
\delta\left(E-E_{n}\left[\{I\}\right]\right).
\end{equation}
For large $\xi$ the typical wells which dominate the DOS
at large negative energies can be considered within a harmonic approximation
by setting
$I(x)  \simeq - I - \frac{1}{2} I_{xx} ~x^2$, with $I_{xx}<0$
so that the resulting spectrum corresponds to a quantum oscillator
with the frequency $\omega=\sqrt{-{I_{xx}}/{m}}$.
The summation over $n$ can be taken to infinity because in the limit
$|s|\rightarrow\infty$ the levels of the harmonic oscillator approach a continuum.
Therefore, for $E<0$ we can
approximate~(\ref{eq:harm}) as
\begin{eqnarray}
\label{eq:dens_max_one_d_mod_}
&&{\nu_{\rm HO}}\left(E\right) \simeq \nonumber \\
&& \sum_{n=0}^{\infty}\langle \delta\left({I_{x}}\right)
\delta\left(-I+\hbar \omega (n +\frac12) -E
\right)
{\Big|}{I_{xx}}{\Big|}\Theta\left(-{I_{xx}}\right)
\rangle. \ \ \ \
\end{eqnarray}
We have computed the statistical average
and carried out the summation over $n$ in Eq.~(\ref{eq:dens_max_one_d_mod_})
by summing over large numbers of states.
The obtained curve is found to be exactly on the top of
the semiclassical curve of Eq.~(\ref{eq:TF_d_less}) shown in
Fig.~\ref{fig:semiclassical}.
Corrections beyond the harmonic approximation and/or
correlations between different points
could be taken into account,
including higher-order derivatives of $I(x)$ at the bottoms of wells.

\begin{figure}
\begin{center}
\includegraphics[width=0.89\columnwidth]{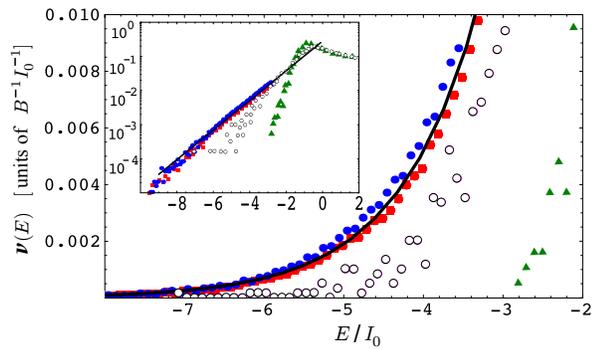}
  \caption{(Color online) The dots represent the numerical
data in linear scale and in logarithmic scale ({inset})
for the low-energy DOS tails in units of $(B I_0)^{-1}$
in a red-detuned speckle potential as a function of $E/I_0$
for different values $s<0$:
$s=-2000$ (red squares), $s=-1000$ (blue circles), $s=-1$ (open circles) and
$s=-0.1$ (green triangles). The black solid line refers to the analytical
semiclassical result~(\ref{eq:TF_d_less}) valid at large $|s|$.
}
\label{fig:red_univ_semic}
\end{center}
\end{figure}

\subsection{Intermediate regime}
Upon decreasing $|s|$ the quasiclassical approximation becomes
inadequate, as shown in Fig.~\ref{fig:red_univ_semic}.
In general, at sufficiently large negative energies $E\ll E_{\xi}$,
the DOS is determined by the deep wells of size
$\lambda\leq \xi$
no matter how
short the correlation length $\xi$ is \cite{john84}.
At higher energies $E\ge E_{\xi}$, the wave function of the
localized state is spread over more than one typical
fluctuation of the disorder potential, that is, $\lambda \geq \xi$.
This picture is supported by the numerical analysis of the spatial extension
of the localized states
in the low-energy tail.
As one can see from Fig.~\ref{fig:eigenstates}
the wave functions for $s=-0.1$
corresponding to the energy range shown in~Fig.~\ref{fig:red_univ_semic}
spread over a region $\lambda\gg \xi$.
The same analysis for $s=-1$ reveals that the typical extension of the wave
function for the low-energy states  is of the same order as $\xi$.
Note that, since the condition $E\ll E_{\xi}$ corresponds to $E/I_0\ll -1/|s|$,
it is not possible to see the transition to the deep localized levels at
$E\ll E_{\xi}$
from the data of~Fig.~\ref{fig:red_univ_semic} for $|s|<1$.

Here we study the low-energy tail of the DOS
for moderate values of $|s|\sim 1$.
This value occurs typically in the experiments with
cold atoms in optical speckles \cite{fallani08}.
In this regime the semiclassical approximation breaks down
due to quantum-mechanical effects  despite that
the interaction between different wells remains negligible.
The decrease of $|s|$ results in a depletion of the DOS
seen in Fig.~\ref{fig:red_univ_semic}. This effect can simply be explained by
the reduction of the number of levels in each potential well.
In the harmonic approximation~(\ref{eq:harm}),
one does indeed find that decreasing  $\xi$ at fixed $B$ leads to
an increase of the distance between levels.
Alternatively, reducing $I_0$ at fixed
$\xi$ causes  a shrinking of the interlevel distance which is
counterbalanced
by a larger decreasing of the wells depth. Thus, the decrease
of $s$ always corresponds to a
reduction of the number of bound states in each well.

In this case we cannot rely on the infinite-sum approximation.
Taking into account only a ground state
in each well severely  underestimates the DOS.
Keeping any constant finite upper limit $M[\{I\}]$ in the sum over levels
remains inadequate due to a quite broad distribution of the number of bound
states in different wells.
A different approach has to be considered. One method is
to extend the semiclassical approximation given by Eq.~(\ref{eq:TF})
in order to include quantum effects. To this end it is useful to
introduce the 1D cumulated DOS~\cite{Lloyd75}
${\cal D}(E, V)=\left({\sqrt{2 m}}/{\pi \hbar}\right)\sqrt{E-V}$
which is related to the total number of states
in a well of size $\lambda$ created at the minimum $V_{\rm min}=-I_{\rm max}$
of the potential $V(x)=-I(x)$,
\[
N(E,V_{\rm min})=\left({\sqrt{2 m}}/{\pi \hbar}\right)\sqrt{E-V_{\rm min}}~\lambda.
\]
The condition to have at least one state
requires that this number is larger than 1, that is,
$0>E\geq V_{\rm min}+\hbar^2/2 m \lambda^2 $. This can be incorporated in
the cumulative DOS by shifting the energy zero~\cite{Lloyd75}:
\begin{equation}
\label{eq:cum_shifted}
{\cal D}(E, V_{\rm min})=\frac{\sqrt{2 m}}{\pi \hbar}\sqrt{E-V_{\rm min}
-\frac{\hbar^2}{2 m \lambda^2}}.
\end{equation}
\begin{figure}
\begin{center}
\includegraphics[width=0.85\columnwidth]{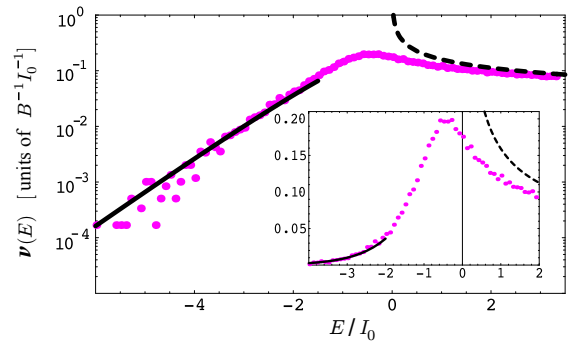}
  \caption{(Color online) The dots represent the DOS for
 a red-detuned speckle potential with $s=-1$  in logarithmic scale
 and  in linear scale   ({inset}).
The solid line describes the
modified semiclassical approximation of
Eqs.~(\ref{eq:spectr_dens_Lloyd})-(\ref{eq:fit}) with $\beta=0.5$
valid in the low-energy tail. At positive energies, the DOS
approaches asymptotically the 1D continuum
$\nu_0(E)=\sqrt{m}/\pi\hbar\sqrt{2~E}$,
indicated by the dashed line.
}
\label{fig:intermediate}
\end{center}
\end{figure}
Changing from the cumulated DOS to the DOS we arrive at
\begin{equation}
\label{eq:spectr_dens_Lloyd}
\nu_{|s|\sim 1} (E,\lambda)=\frac{\sqrt{m}}{\sqrt{2}\pi \hbar}\int_{-\infty}^{E-\frac{\hbar^2}{2 m \lambda^2}}
\frac{{\tilde P}[-V]}{\sqrt{E-V-\frac{\hbar^2}{2 m \lambda^2}}}
~d V,
\end{equation}
where ${\tilde P}[-I]$ is the normalized probability distribution
function of  minima of the red-detuned speckle potential, so that
${\tilde P}[I]$ is the similar distribution of maxima of the
blue-detuned speckle:
\begin{equation}
\label{eq:prob_min}
{\tilde P}[I]=\frac{n_{\rm max}(I,\xi,\kappa)}
{\int_0^{\infty} n_{\rm max}(I',\xi,\kappa)~d{I^{'}}}.
\end{equation}
Moreover, because we know that the extension of the wave function of these states
does not exceed the typical size of the potential wells,
we can take the energy of the localized state proportional to the zero-point energy
inside the well, that is,
\begin{equation}
\label{eq:fit}
{\hbar^2}/{2 m \lambda^2}\sim \beta~|E|,
\end{equation}
where the parameter $\beta$ cannot be determined exactly within this method.
However the result is not extremely sensitive to the choice of $\beta$
in the limits $0<\beta<1$.
Choosing the value $\beta\sim 0.5$ the modified
semiclassical method based on Eqs.~(\ref{eq:spectr_dens_Lloyd}) and (\ref{eq:prob_min})
reproduces very well the data for $s=-1$,
as shown in Fig.~\ref{fig:intermediate}.
We also tried to fit the computed DOS by a stretched exponential.
We found that it cannot be done well in the whole window of numerical data. However,
in smaller windows it can be fitted by a stretch exponential with the exponent
varying between 1 and 1.5 and weakly depending on $\beta$.

\subsection{Quantum limit}

When the correlation length of the disorder $\xi$ is reduced to
$|s|\leq 1$, the wave function of the localized states in the low-energy tail
$-1/|s|<E/I_0<0$
is spread over a distance $\lambda$ larger than the typical distance
between two minima (see Fig. \ref{fig:eigenstates}).
Therefore, the probability of constructing
a local wave function is not directly related to the probability distribution
of the bare potential $P[I]$ or that of its extrema  $\tilde{P}[I]$.
The interplay between different wells has to be taken into account as well.
This can be done by considering
the probability $P_{\lambda}[I]$
of the potential integrated over the characteristic extension $\lambda(E)$
of the state of energy $E$
\cite{Lifshitz68}.

For finite $\lambda(E)> \xi$ we can divide the region occupied by the wave function
into $n= \lambda(E)/ \xi$ parts, which are approximately uncorrelated. Assuming
that the correlation inside each part is strong enough
we can approximately replace the distribution of potential
integrated over the correlation length $P_{\xi}[I]$ by the bare
distribution (\ref{eq:ray}).
Then the distribution  integrated  over the width of wave function
$P_{\lambda}[I]$  can be computed using the characteristic function method.
The  characteristic function is defined as $ f(t)= \langle e^{i t I}\rangle $
so that $f_{\lambda}(t)=[f_{\xi}(t/n)]^n$. This gives
\begin{equation}
\label{eq:fin-n}
P_{\lambda}[I] = \frac1{I_0} \frac{n^n}{\Gamma(n)}
\left(\frac{I}{I_0}\right)^{n-1} e^{-n I/I_0},
\end{equation}
with $n= \lambda(E)/ \xi$.
It can be shown
that for very large $\lambda$ the integrated exponential
distribution~(\ref{eq:fin-n}) approaches asymptotically a Gaussian law:
\begin{equation}
\label{eq:Gauss}
P_{\lambda}[I]\stackrel{\lambda\gg 1}{\sim}\frac{\sqrt\lambda}{\sqrt{2\pi}I_0\sqrt{\xi}}
\exp{\left(-\frac{\lambda(I-I_0)^2}{2I_0^2\xi}\right)}.
\end{equation}
By reversing the argument, we can say that for a given energy there is an
optimal well of size $\lambda$
created by the average potential $V$, which may bind a state whose
zero-point energy should not
be larger than $\hbar/2 m\lambda^2$.
Therefore, the negative low-energy tail of the DOS can be calculated
with a variational principle from~\cite{Lloyd75}
\begin{equation}
\label{eq:spectr_dens_Lifs}
\nu_{|s|\ll 1} (E,\lambda)\sim\frac{\sqrt{m}}{\sqrt{2}\pi \hbar}
\int_{-\infty}^{E-\frac{\hbar^2}{2 m \lambda^2}}
\frac{{P}_{\lambda}[-V]}{\sqrt{E-V-\frac{\hbar^2}{2 m \lambda^2}}}
~d V.
\end{equation}
By making the variation with respect to $\lambda$, for large negative
energies $|E|\gg I_0$,
we find that the DOS of Eq.~(\ref{eq:spectr_dens_Lifs})
goes asymptotically as
\begin{equation}
\label{eq:spectr_dens_Lifs_bis}
\nu_{|s|\ll 1} (E)\sim \exp{\left(
-\frac{8\sqrt{3}}{9 \sqrt{2}}\frac{\hbar|E|^{3/2}}{\sqrt{m} \xi I_0^2}
\right)}.
\end{equation}
The limit $s\rightarrow 0^-$ of very short disorder correlation
length corresponds thus
to the white-noise potential
$$
\langle V(x) V(x')\rangle_{P_{\lambda}}=\kappa_0^2 \delta(x-x'),
$$
where $\kappa_0^2\equiv I_0^2 \xi$ has to remain finite as
$\xi\rightarrow 0$. Therefore, we see that the asymptotic 
DOS~(\ref{eq:spectr_dens_Lifs_bis})
reproduces the typical Lifshitz tail law for unbounded delta-correlated potential
 \begin{equation}
\label{eq:spectr_dens_Lifs_tris}
\nu_{|s|\ll 1} (E)\sim \exp{\left(
-\frac{2\sqrt{3}}{9}\left(\frac{|E|}{{ E}_{\rm L}}\right)^{3/2}
\right)},
\end{equation}
where ${ E}_{\rm L}=\left(m \kappa_0^4/\hbar^2\right)^{1/3}/2$ is nothing but
the 1D Larkin energy~\cite{Falco08,Falco09}.

\begin{figure}
\begin{center}
\includegraphics[width=0.98\columnwidth]{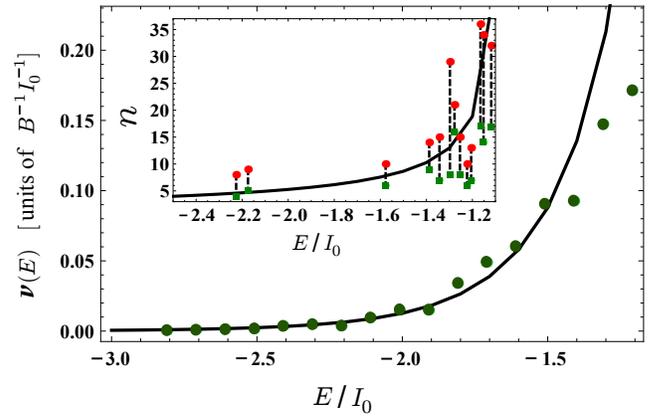}
  \caption{(Color online) The dots show the low-energy DOS tail computed
  numerically for a particle in red-detuned speckle potential
with  $s=-0.1$.
The black solid line corresponds to the variational
result of Eq.~(\ref{eq:spectr_dens_Lifs}) with  $P_{\lambda}[I]$ given by
Eq.~(\ref{eq:fin-n}). The inset shows the number of wells occupied by the
states as a function of energy. The green squares and red circles
are computed for the first few states in a particular speckle
sample using the attenuation factor of 10 and 100, respectively.
The solid line is the prediction of the variational
calculations.}
\label{fig:reds01}
\end{center}
\end{figure}

In Fig.~\ref{fig:red_univ_semic} we  show some numerical results
for the systems with values of $s$  down to $s=-0.1$.
For such a small value of $|s|$, the typical wave-functions of the states
in the low-energy tail are spread over a distance a few times larger than the
typical distance between neighboring wells.
These states are not yet sufficiently extended in order to
reach for the distribution
of $P_{\lambda}[I]$
the asymptotic Gaussian probability shape of Eq.~(\ref{eq:Gauss}).
The probability distribution $P_{\lambda}[I]$ given by Eq.~(\ref{eq:fin-n})
lies somewhere between  the exponential
law~(\ref{eq:ray}) and the Gaussian curve of Eq.~(\ref{eq:Gauss}).
To compute the DOS for moderately small $s$, we substitute  Eq.~(\ref{eq:fin-n})
into Eq.~(\ref{eq:spectr_dens_Lifs}) and minimize it with respect
to the number of occupied wells $n=\lambda/\xi$.
This  gives the result which  interpolates between the semiclassical
result of Eq.~(\ref{eq:spectr_dens_Lloyd})
and the Lifshitz tail~(\ref{eq:spectr_dens_Lifs_tris}).
The corresponding DOS
computed for $s=-0.1$ is shown in Fig. \ref{fig:reds01} together with numerical
data. The variational prediction for the number of occupied states is shown in the 
inset.
The dots in the inset depict the number of occupied wells for the first few states
in a particular speckle sample.  To estimate this number from the numerical data
we calculate the width of each wave function using the criterion of the amplitude
attenuation  by factors of 10 and 100. We find a good agreement with the results of
the variational calculation.

\begin{figure}
\includegraphics[width=0.9\columnwidth]{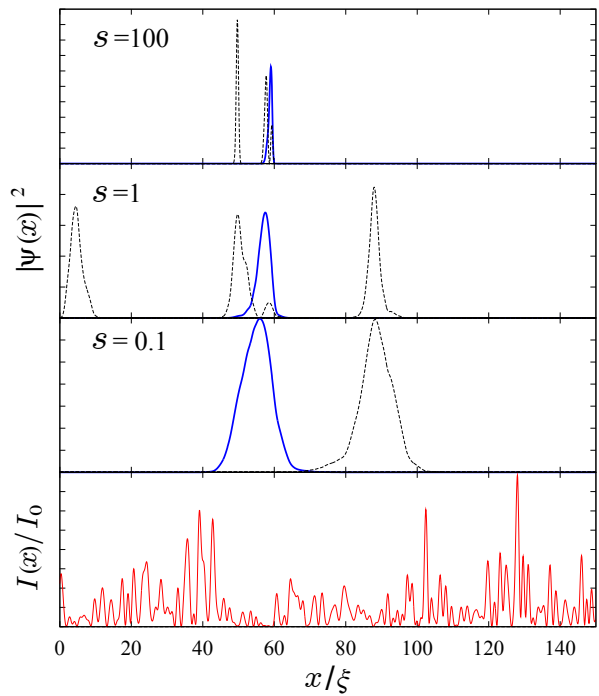}
  \caption{(Color online) Probability density distribution of the lowest lying
eigenstates of a \textit{blue-detuned} speckle pattern (in the bottom panel),
for three different intensities with $s=0.1,\ 1,\ 100$.
Only a window of length $150\xi$ of the whole system is shown.
The figure shows the ground state (blue solid line) and several first
excited states (black dashed line).
}
  \label{fig:eigenstates-blue}
\end{figure}

\section{DOS: blue-detuned speckle}
\label{sec:blue}

The random potential created by a blue-detuned laser speckle is bounded
from below so that the DOS vanishes at energies
lower than the natural boundary $E_0=0$.
Analogously to the red-detuned problem, the localized region of the spectrum
is controlled by the parameter $s$. The lowest-lying
eigenstates for three different intensities of the same speckle profile,
corresponding to $s=0.1, 1$, and $100$ are depicted in
Fig.~\ref{fig:eigenstates-blue}.
This picture shows that with decreasing $s$  the wave functions extend over
larger numbers of potential wells.

\begin{figure}
\begin{center}
\includegraphics[width=0.9\columnwidth]{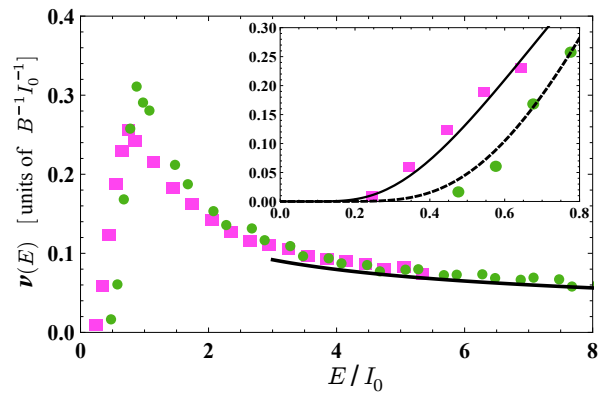}
\end{center}
  \caption{(Color online) The one particle DOS
for  a blue-detuned speckle potential computed numerically  for
$s=0.1$ (greed circles) and $s=1$ (magenta squares).
The black solid line is the DOS of free particle.
{Inset.} The best simultaneous fit of the Lifshitz tail of both
DOS by Eq.~(\ref{lifshitz-1}) with the same $c_0=0.8141$ and $A=0.4098$.
}
\label{fig:blue-small-s}\end{figure}

\subsection{Small $s$ limit}

The DOS computed numerically for  $s=0.1$ and $s=1$
are shown in Fig.~\ref{fig:blue-small-s}. At large energies the DOS approaches
the limit of free particle while at small energies it exhibits the Lifshitz's
tail behavior. It is well known that the DOS near the bottom of a
bounded-from-below disorder potential is controlled by the existence of
very large regions ``free" of
disorder potential.  The probability of their appearance
in the speckle pattern can be estimated as follows.

We assume the existence of a bound state with energy $E$
localized inside this region.
In general, this
region has the average intensity that is not exactly zero but well
below the energy level $E$. Since we are interested in the asymptotic
low energy behavior, we can
fix the largest intensity to $E\ll I_0$.
However, very small intensity corresponds also to a vanishing electric field.
The probability to have a region of size $L$ with the electric field in the
interval between ${\cal E}(x)$ and ${\cal E}(x)+d{\cal E}(x) $  is given by
\begin{eqnarray}\label{W-1}
&&\!\!\!\!\!\!\!\!\! d W[{\cal E}(x),d{\cal E}(x)] \nonumber \\
&& = \exp \left[- \int\limits_0^{L} {\cal E}^{*}(x) G^{-1}(x-x'){\cal E}(x')dx dx'
       \right] \nonumber \\
&& \times (\det G)^{-1}\prod\limits_{x\in[0,L]}
\frac{d{\cal E}(x) d{\cal E}^{*}(x)}{\pi a_0},
\end{eqnarray}
where $a_0$ is the UV cutoff and the inverse correlator is defined by
\begin{eqnarray}
\int dx'' G(x-x'')G^{-1}(x''-x') = \delta(x-x').
\end{eqnarray}
The  Fourier transform
of the electric field correlator (\ref{eq:sinc_one_d}) reads
\begin{equation} \label{G-fourier}
\tilde{G}(k) = I_0 /D\, \Theta({\pi}\, D-|k|),
\end{equation}
where $\Theta(x)$ is the Heaviside step function. In what follows, we  use
$D \sim 1/\xi$ absorbing the numerical factor of $0.88$ into the overall
coefficient. For small $s$  the eigenstates
spread over distance $L$ much larger than the typical well width $\xi$.
In this regime  $\det G$ can be estimated using Eq.~(\ref{G-fourier})
as follows:
\begin{equation}
\det G = \prod\limits_{k=-\pi/\xi}^{-\pi/L} I_0 \xi
  \prod\limits_{k=\pi/L}^{\pi/\xi} I_0 \xi   = (I_0 \xi)^{2L/\xi}.
\end{equation}
Substituting this into Eq.~(\ref{W-1}) and putting $\cE(x)=0$ in the exponential
gives the probability for the region of size $L$ to have the electric field
in the interval between $0$ and $d\cE(x)=E^{1/2}e^{i \theta}$,
where $\theta$ is arbitrary. Integrating out $\theta$ and summing over $x$
we obtain
\begin{equation}\label{W0-2}
W[0,E^{1/2}] \sim (I_0 \xi)^{-2L/\xi} (2 E /a_0)^{2L/a_0}.
\end{equation}
Since we are interested in the leading
exponential behavior, we can replace in  Eq.~(\ref{W0-2})
$a_0$ by $\xi$ and arrive at
\begin{equation} \label{W0-3}
W[0,E^{1/2}] \sim \exp \left[-\frac{2L}{\xi}\ln\left(\frac{I_0}{E}
\right)\right].
\end{equation}
The region being almost free of disorder has a bound state with energy
$E=c_0^2\hbar^2/(2m L^2)$, where $c_0$ is a constant of order $1$.
Using the latter in Eq.~(\ref{W0-3}) to express $L$ in terms of $E$,  we obtain
the Lifshitz's tail for the energy distribution as
\begin{eqnarray} \label{lifshitz-1}
\nu(E) \approx A \exp\left[-c_0 s^{-1/2}\sqrt{\frac{I_0}{E}}
\ln\left(\frac{I_0}{E}\right)\right].
\end{eqnarray}
The logarithmic corrections to the Lifshitz tail of the same form as in
Eq.~(\ref{lifshitz-1}) was found for the uncorrelated disorder by mapping
the problem
to the random walk of a classical particle in a medium with traps~\cite{politi88}
and using a variational approach
\cite{luck89}. We fit using Eq.~(\ref{lifshitz-1}) the lower energy tail
of the DOS  computed numerically
for $s=0.1$ and $s=1$. The corresponding curves  are shown in the inset
of Fig.~\ref{fig:blue-small-s}. We find that taking into account
the logarithmic corrections is necessary in order to fit the data
in the energy interval that is accessible numerically.

\subsection{Large $s$ limit}

\begin{figure}
\begin{center}
\includegraphics[width=0.9\columnwidth]{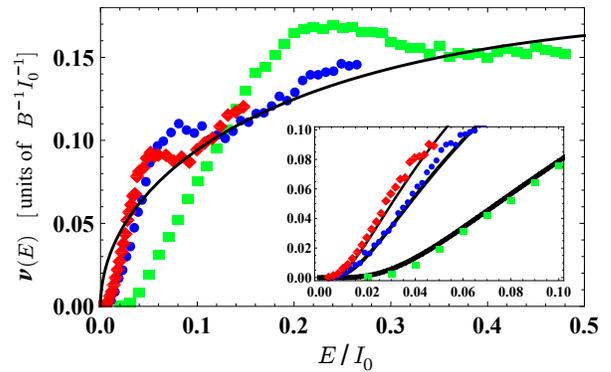}
\end{center}
  \caption{(Color online) The DOS of a particle
in a blue-detuned speckle potential are computed numerically  for
$s=2000$ (red diamond), $s=1000$ (blue circles) and $s=100$ (green squares).
The black solid line refers to the analytical semiclassical
result~(\ref{eq:TF_d_less_blue}) valid at large $s$
and $E/{I_0}\gtrsim1/\sqrt{s}$.
{Inset.} The best simultaneous fit of the Lifshitz tail of three
DOS by $\nu(E)\approx A \exp\left[-c_0 s^{-1/4} \sqrt{I_0/E} \ln (I_0/E) \right]$
with the same $c_0=0.6375$ and $A=0.3425$.
}
\label{fig:blue-large-s}\end{figure}

The DOS computed numerically for  $s=100, 1000$, and $2000$
are shown in Fig.~\ref{fig:blue-large-s}. The DOS exhibits a quite
flat crossover region between the narrow Lifshitz's tail and the
large energy asymptotic behavior corresponding to the free particle.
For general $s$ the Lifshitz tail of the DOS is expected to
have the form
\[
\nu(E) \sim \exp\left[- c_0 \sqrt{{I_0}/{E}}\,
{\cal F}\left(s, \ln\left({I_0}/{E}\right) \right)
\right],
\]
where ${\cal F}(s,t)$ is a universal function depending only on the form of
$G(x)$. Fitting the DOS computed numerically 
(see inset of Fig.~\ref{fig:blue-large-s}) we find that for  $s\in[100,2000]$
this function can be approximated by  ${\cal F} (s,t) \approx s^{-1/4}\, t$.
Thus, the correlations in the speckle potentials growing with $s$
renormalize the effective exponent of $s$ in the exponential of the Lifshitz
tail.

To compute the DOS in the crossover regime, we use the
semiclassical approximation and obtain
\begin{align}
\label{eq:TF_d_less_blue}
\nu_{s\rightarrow \infty} (E)=&\frac{1}{I_0}\int_{0}^{E} \nu_0(E-I) ~e^{-I/I_0}~d I\nonumber\\
=&\sqrt{\frac{m}{2 \pi\hbar^{2}I_{0}}} e^{-E/I_0}
{\rm erfi}\left[\sqrt{\frac{E}{I_0}}\right],
\end{align}
where ${\rm erfi}(x)$ is the imaginary error function.
The DOS~(\ref{eq:TF_d_less_blue}) goes asymptotic
as $\sim \nu_0 (E)$ for large positive energies. Near the boundary of
the spectrum $E\rightarrow 0$ it behaves as
\[
\nu_{s\rightarrow \infty} (E)
=
\frac{1}{\pi}\sqrt{\frac{2m}{\hbar^{2}I_{0}}}\sqrt{\frac{E}{I_0}}.
\]
The comparison between Eq.~(\ref{eq:TF_d_less_blue}) and the numerical data
data in Fig.~\ref{fig:blue-large-s} shows that there is a range of energies
where the two results agree quite well.

\subsection{Effective mobility edge}

The possible existence of the effective mobility edge in 1D systems
with certain correlated  disorder has been considered in several
works~\cite{izrailev99,lugan09}.
It is expected, for example, for the potentials with correlations
having a finite support in Fourier space. In particular as follows from
Eq.~(\ref{G-fourier})
the speckle potential has no Fourier components with $|k|>\pi/D$. As a result,
in the Born approximation there is no backscattering for particles with momentum
larger than $\hbar k >  \hbar \pi/D$ that defines the effective mobility edge.
The localization length above this edge is determined by corrections to the Born
approximations and larger by one order of magnitude. In our unities the effective
mobility edge is expected at
\begin{equation}
E_{\rm M}=I_0[1+(0.88\pi)^2/s],
\end{equation}
which gives $E_{\rm M}/I_0 \approx 8.64$ for $s=1$  and  $E_{\rm M}/I_0 \approx 1.008$
for $s=1000$. At these energies the DOS is already
very close to the DOS of a free particle. One cannot see any signature of the
effective  mobility edge like it happens with the DOS and the real mobility edge
in higher dimensions. To probe the effective mobility edge, one has to study the
asymptotic behavior of the wave functions at this energy~\cite{sanchez07}
which is beyond of the scope of this article.


\section{Summary}
\label{sec:sum}

In this work we have studied the DOS of a quantum particle in
a laser-speckle potential using the statistics of the speckle potential
profile and numerical simulations. The DOS
in the speckle potential can be characterized by a single parameter $s$,
which depends on the correlation length of the disorder, the average value
of the laser intensity and the mass of the particles.
We have considered both cases of red-detuned and blue-detuned potentials for which
the random potential is bounded from
above and from below, respectively.

For the negative-energy region of DOS in the
\textit{red-detuned} speckle
there are three different regimes: semiclassical ($|s|\gg1$),
intermediate ($|s|\sim1$), and quantum ($|s|<1$).
For strongly correlated disorder, $|s|\gg1$,
the tail of the DOS is related
to the statistical distribution of the deep minima of the disorder potential.
In the quantum limit with $|s|<1$,
the DOS is controlled by smoothing effects leading to the Lifshitz tail
of uncorrelated Gaussian  potential.
In the crossover regime, $|s|\sim1$, the DOS can be described within
the semiclassical approach by including quantum corrections.

In the case of a \textit{blue-detuned} speckle potential,
the DOS near the bottom of the spectrum is given
by the Lifshitz tail related to the existence of
large regions with negligible intensity.
We have found  logarithmic corrections to the celebrated Lifshitz
result due to the continuous nature of the distribution of the potential.
The relevance of these corrections for typical setup parameters
in current experiments
is supported by our numerical results.

\acknowledgments{
We  thank
Cord M\"uller, Valery Pokrovsky, Thomas Nattermann, Henk Stoof and
Rembert Duine for useful discussions.
J. G. thanks the Physics Department of the University of Florence and
LENS for hospitality. }


\appendix*

\section{}

\noindent Because $E_x$ has nonzero correlations only with itself, the total
joint probability in Eq.~(\ref{eq:dens_min_one_d}) can be written as
\begin{align}
\label{eq:joint_prob_one_d_total}
p(E,E_x,E_{xx})=p(E_x)p(E,E_{xx}).
\end{align}
In rescaled unities of $I_0$ and $\tilde\xi$ we can immediately write
\begin{align}
p(E_x)\equiv{(\pi^2\langle E_x^{\ast}E_x\rangle)^{-1/2}}e^{-\frac{|E_x|^2}{\langle E_x^{\ast}E_x\rangle^2}}
={\pi^{-1}}e^{-{|E_x|^2}}.
\end{align}
The joint probability for the two correlated variables $E$ and $E_{xx}$
can be written using matrix representation as
\begin{align}
\label{eq:joint_prob_one_d}
p(E,E_{xx})=
\left(\pi^2{\rm det}~{\bM}\right)^{-1}
e^{-{\bA}^{\dagger}\bM^{-1} \bA},
\end{align}
with $\bA^{\dagger} = (E^*,E^*_{xx})$
and
\begin{align}
\bM\equiv\left(\begin{array}{cc} \langle E E^{\ast}\rangle & \langle E E_{xx}^{\ast}\rangle\\
\langle  E_{xx} E^{\ast}\rangle & \langle E_{xx} E_{xx}^{\ast}\rangle
\end{array}\right)=
\left(\begin{array}{cc} 1 & -1\\
-1 & \kappa^{-1}
\end{array}\right),
\end{align}
which has an inverse matrix that reads
\begin{align}
\bM^{-1}=\frac{1}{\kappa^{-1}-1}
\left(\begin{array}{cc} \kappa^{-1} & 1\\
1 & 1
\end{array}\right).
\end{align}
Equation~(\ref{eq:joint_prob_one_d}) can be rewritten as
\begin{align}
\label{eq:joint_prob_one_d_bis}
p(E,E_{xx})=\frac{\pi^{-2}}{\kappa^{-1}-1} e^{-\frac{1}{\kappa^{-1}-1}
\left(
|E|^2 \kappa^{-1}+|E_{xx}|^2+2 {{\cal R}e\left[E^{\ast} E_{xx}\right]}
\right)}.\nonumber
\end{align}
Using the following variable transformation
$E(x)\equiv \sqrt{I(x)}~e^{i \theta(x)}$ and setting $I_x=0$, we obtain
\begin{align}
E_x=& i \theta_x I^{1/2}e^{i \theta(x)}\\
E_{xx}= & \left(
\frac{I_{xx}}{2 I^{1/2}}-I^{1/2}\theta_x^2
\right)e^{i \theta}+i \theta_{xx}I^{1/2}e^{i \theta}.
\end{align}
Therefore, the joint probability~(\ref{eq:joint_prob_one_d_total})
can be written as
\begin{align}
\label{eq:joint_prob_one_d_totalbis}
p(E,E_x,E_{xx})=J_0~ p(\theta,\theta_x,\theta_{xx},I,I_x=0,I_{xx}),
\end{align}
where
\begin{align}
\label{eq:joint_prob_one_d_aux}
p&(\theta,\theta_x,\theta_{xx},I,I_x=0,I_{xx})=
\frac{\pi^{-3}}{\kappa^{-1}-1}\\
&e^{-I \theta_x^2
-\frac{1}{\kappa^{-1}-1}\left[
\kappa^{-1} I+
\left(
\frac{I_{xx}}{2 I^{1/2}}-I^{1/2}\theta_x^2
\right)^2
+I \theta_{xx}^2+I_{xx}-2 I\theta_x^2
\right]
},\nonumber
\end{align}
and the Jacobian is $J_0=2^{-3}$.
Integration over $\theta$ yields a factor $2\pi$. We can also analytically
integrate out $I$ and $\theta_{xx}$, which is a Gaussian integral. As a result,
we arrive at Eq.~(\ref{eq:rhomin}).



\begin{thebibliography}{99}
\bibitem{numerical}
We use the following prescription:
${H}_{ij} = {1}/{\Delta\tilde{x}^2}\left(2\delta_{ij}%
-\delta_{i+1,j}-\delta_{i-1,j}\right) + s \tilde{I}(j\Delta \tilde{x})%
\delta_{ij}$, with vanishing boundary conditions, $\psi_0=\psi_{N+1}=0$.
The number of grid points is 8192.
\end{thebibliography}
\end{document}